\documentclass{article}
\usepackage{graphicx}
\usepackage{subcaption}

\begin{document}
\title{Malware classification using transfer learning}
\author{Hikmat Farhat and Veronica Rammouz}
\date{\small{Notre Dame University-Louaize}}
\maketitle
\begin{abstract}
With the rapid growth of the number of devices on the Internet, malware poses a threat not only to the affected devices but also their ability to use said devices to launch attacks on the Internet ecosystem. Rapid malware classification is an important tools to combat that threat. One of the successful approaches to classification is based on malware images and deep learning. While many deep learning architectures are very accurate they usually take a long time to train. In this work we perform experiments on multiple well known, pre-trained, deep network architectures in the context of transfer learning. We show that almost all them classify malware accurately with a very short training period.
\end{abstract}
\section{Introduction}

Malicious software (malware) are computer programs that compromise hosts for various reasons. They could take control of the host for ransom (ransomware) or they could be used for launching attacks, typically Denial of Service (DoS), against other hosts and networks. The latter type depends on a large number of hosts being compromised, but  the increase in the popularity of IoT devices has made the task much easier \cite{mirai}.

\noindent Traditional malware detection and anti-virus systems use signature-based method. These methods fail in the presence of polymorphic or mutable code. Furthermore, a large number of malware seen on the internet are small variations of few known malware where the difference between the "new" and "old" malware being as low as 2\% \cite{commACM} and signature-based method fail to detect those also.

\noindent Recently, machine learning methods were used to classify malware because of their accuracy in detecting and classifying similar patterns, such is the case in malware variants. One such approach converts binary files into grayscale images and uses Deep Convolution Neural Networks (DCNN) which are known to be very successful in image classification and object detection.

\noindent However, the training of DCNN can take weeks\cite{vgg16}. Therefore, it is important to find a solution to speed up the training. One such solution is the use of transfer learning. By using known pre-trained networks, one could cut the training time drastically. In this paper we investigate the efficacy of transfer learning for malware classification. We do this by performing experiments on four widely used DCNN for image classification: ResNet50 and Resnet152 \cite{resnet}, MobileNet\cite{howard_mobilenets_2017}, and VGG16\cite{vgg16} on the same dataset.

\section{Related Work}
 Malware classification methods can be divided into dynamic and static. Dynamic methods execute the malware and observe its behavior. Static methods extract features from the malware without actually executing it.  During the last decade many machine learning methods for malware classification were proposed. 
 \noindent The work in \cite{yuxin_malware_2019} uses deep belief networks. Support Vector Machines and kNN are used for malware classification in \cite{narayanan2016performance}.
 
 \noindent Entropy is a measure of disorder whereby encrypted binary have a higher entropy. Methods based on entropy have been used for malware detection \cite{bat2017entropy}\cite{lyda2007using}\cite{han2015malware}.

 \noindent first to propose to use image classification techniques based on the grayscale image representation of malware is \cite{nataraj_malware_2011}. In similar approach, but using transfer learning  \cite{rezende_malicious_2017} use ResNet50 for classification of the Microsoft Big Challenge. Also, a variation of VGG16\cite{vgg16} is used in \cite{kalash_malware_2018} for malware classification.

 \noindent All the above mentioned methods obtained a classification accuracy between 90-99\%. Rather than optimize for very high accuracy our interest is in the training time of of the networks. In particular, to investigate which of the well known network architectures can be trained for a few epochs, as opposed to hundreds, and yet obtain an accuracy above 95\%.

\section{Dataset and Methodology}
In this paper we use the Microsoft Malware Classification Challenge dataset \cite{ronen_microsoft_2018}. The dataset contains 10868 labeled samples from 9 malware families with two files associated with each sample: a binary file ('.byte') and assembly file ('.asm'). The distribution of the samples on the 9 different families is shown in Table 1.
\begin{table}
    \begin{center}
    \begin{tabular}{|c|c|c|}
        \hline
        Family & Samples & Type \\
        \hline
        Gatak & 1013 &Backdoor\\
        Kellihos\_ver1 & 398 & Backdoor\\
        Kelihos\_ver3 & 2942 & Backdoor\\
        Lollipop & 2478 & Adware \\
        Obfuscator & 1228 & Obfuscated Malware\\
        Ramnit & 1541 & Worm\\
        Simda & 42 & Backdoor\\
        Tracur & 751 & Trojan\\
        Vundo & 475 & Trojan\\
        \hline        
    \end{tabular}
\end{center}
\label{tab:1}
\caption{Microsoft Challenge Dataset}
\end{table}

\begin{figure}[ht]
      \includegraphics[width=.3\linewidth]{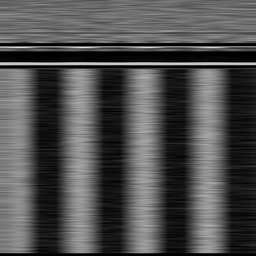}  
      \includegraphics[width=.3\linewidth]{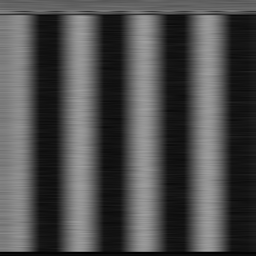}  
        \includegraphics[width=.3\linewidth]{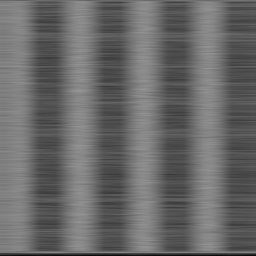}      
\caption{Three samples from the Gatak family converted to images.}
\label{fig:1}
\end{figure}

\noindent Following  \cite{nataraj_malware_2011} we convert binary files to grayscale image that serve as an input to the neural network classifiers.
Towards that end, each binary sample ('.byte') is converted into a grayscale image as follows. Each file contains 16 hex numbers per line where each byte was considered as a grayscale pixel value in the range 0-255. Then the images were resized to dimension (256,256). In  Fig.\ref{fig:1} we show three such samples from the Gatak family.

\noindent The preprocessed images are used as input for a neural network with two connected parts, one after the other. The first part is one of the well known architectures pre-trained on the ImageNet dataset with the last, classification layer removed. The second part consists of a fully connected layer with 1024 neurons followed by a 9 node softmax classification layer.

\begin{figure}[b!]
    \centering
    \includegraphics[width=0.9\textwidth]{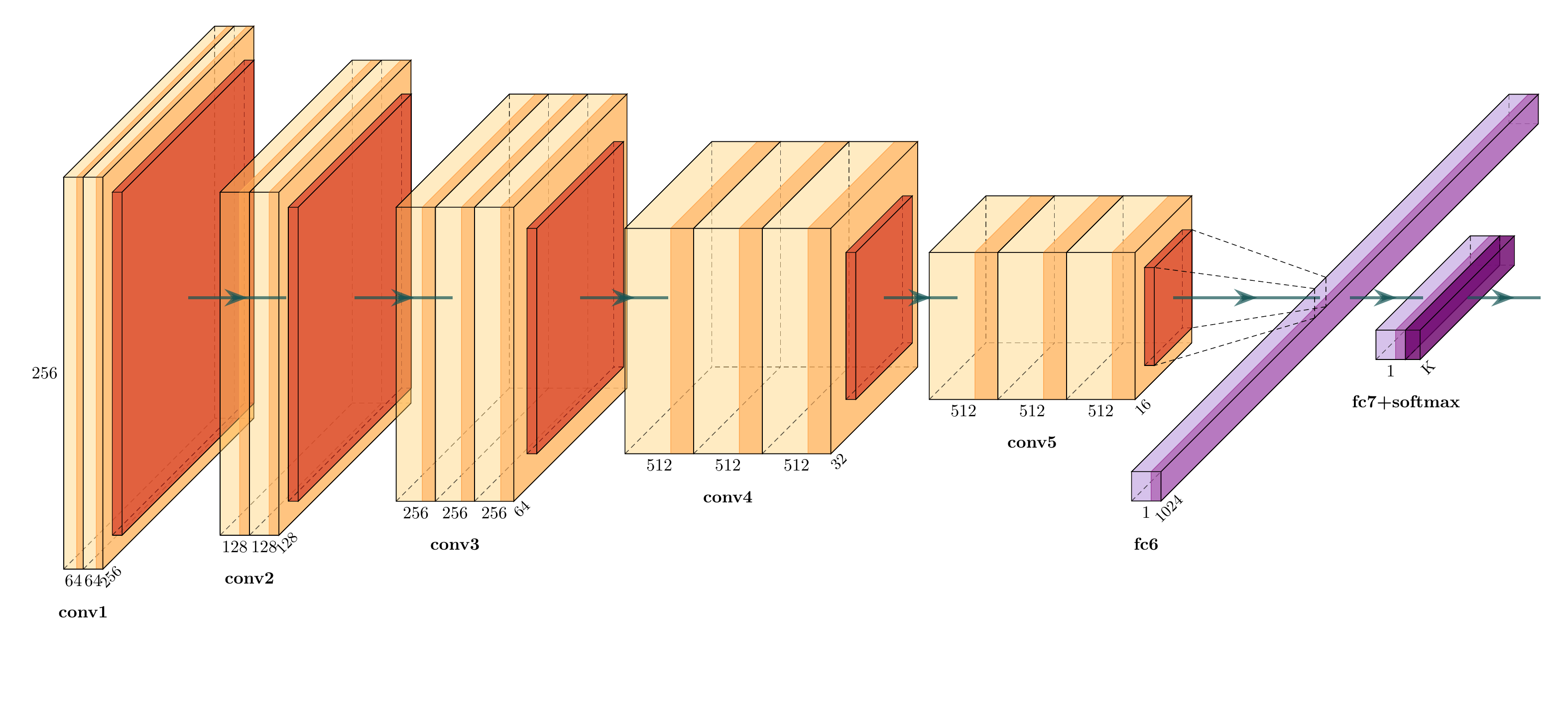}
    \caption{Deep convolution network trained from random weights to compare to pre-trained networks}
    \label{fig:nn}
\end{figure}
\noindent Training is also done in two phases. In the first phase, which lasts for 15 epochs, the pre-trained network is \textit{frozen} and only the added two layers are optimized. The second, fine tuning phase with 10 epochs, continues where the first left off,  the pre-trained network is \textit{unfrozen} and the complete network is trained with a very small learning rate (10e-5).

\noindent Since our goal is to assess the viability of tranfer learning not just a single architecture we have performed the above training with the following list of pre-trained networks.
\begin{enumerate}
    \item VGG16
    \item Resnet50
    \item Resnet152
    \item MobileNet
\end{enumerate}
\noindent All the above architectures were pre-trained on the ImageNet dataset \cite{imagenet}. Furthermore, to gauge the time saving component of transfer learning we ran the same experiments using a deep convolution neural network (CNN) shown in Fig. \ref{fig:nn}.

\noindent The CNN contains 13 convolution and 5 max-pooling layers and the last two layers are fully connected. The activation function for all layers except the last, which uses a softmax activation, are ReLU.
In Fig. \ref{fig:nn} the yellow boxes are convolution layers and the orange ones are max-pooling layers. 

\begin{figure}[t]
    \centering
    \includegraphics[width=0.8\textwidth]{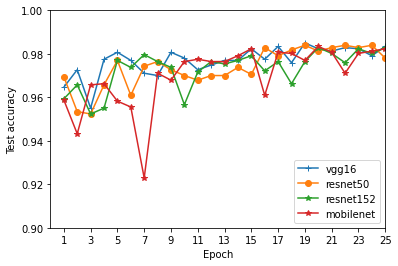}
    \caption{Test accuracy for the various pre-trained networks}
    \label{fig:2}
\end{figure}

\noindent The experiments were performed on Kaggle which uses a NVIDIA Tesla P100 GPU with 16GB or RAM. The code was written in Python using Tensorflow/Keras. The 10686 samples were randomly divided into 9000 for training and 1868
for testing. The training was performed using the Adam optimizer \cite{kingma_adam_2017}, with a learning rate of 0.01, and a batch size of 32 samples. The results for the two phase transfer learning described above are shown in Fig. \ref{fig:2}. 
\begin{figure}[t]
    \centering
    \includegraphics[width=0.6\textwidth]{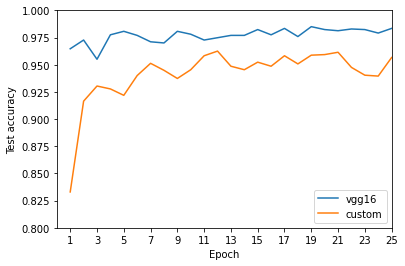}
    \caption{Comparison between pre-trained vgg16 and CNN}
    \label{fig:3}
\end{figure}
\section{Results}
\noindent The accuracy of the test data set for the different networks is shown in Fig. \ref{fig:2}. As can be seen from the figure all the architectures gave similar results. Furthermore, the results are excellent considering the short training duration and there was no hyper-parameter optimization was performed. Also, after just a single epoch all of them give an accuracy above 94\%.

\noindent To illustrate the time saving feature of transfer learning we conducted the same experiment using the convolution network shown in Fig. \ref{fig:nn}. To compare with the pre-trained networks the training was done for 25 epochs using the same hyper-parameters.  A comparison of the test accuracy between vgg16  are shown in Fig. \ref{fig:3}. Note that the network is very similar to vgg16 and needs to be run for more than 75 epochs to reach a comparable accuracy of what the pre-trained model can obtained in 25 epochs.

\begin{figure}[t]
    \centering
    \includegraphics[width=0.8\textwidth]{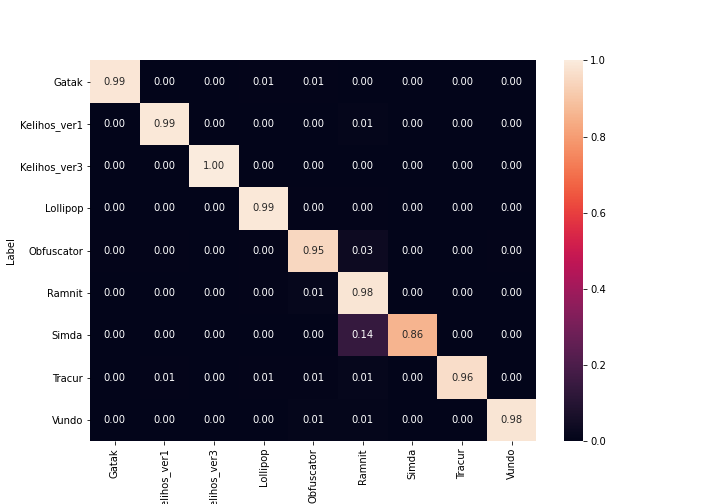}
    \caption{Confusion matrix for ResNet152}
    \label{fig:4}
\end{figure}
\noindent The obtained accuracy does not reflect the prediction accuracy of individual malware families. Toward that end we have computed the confusion matrix which is 
shown Fig. \ref{fig:4}  for ResNet152. All the others are similar. In particular, the low accuracy of the prediction of Simda is mostly due to the very small number of samples.

\section{Conclusion}
\noindent In this work we investigated the efficacy of transfer learning for malware classification. Toward that end, we have performed experiments on four pre-trained networks for the purpose of classifying malware. In particular the classification was performed on the Microsoft Classification Challenged dataset which were converted to grayscale images. All network architectures gave more than 95\% accuracy using very few training epochs. This is very promising since they were trained on ImageMet. This shows that transfer learning is reliable since all the different networks gave more or less the same behavior.

The question of loss of information due to the resizing of the images is worth investigating \cite{yuan_byte-level_2020}. Another aspect worth investigating which we leave to future work, is the inference time of the studied network architectures on small computer-on-chip devices such as the NVIDIA Jetson Nano.
\bibliographystyle{abbrv}
\bibliography{malware.bib}
\end{document}